\documentclass[useAMS,usenatbib,usegraphicx]{mn2e}

\newcommand\degree{$^{o}$}
\newcommand\kms{km s$^{-1}$}
\newcommand\etal{et al.}

\title[ 53 MHz OH line ]
{A search for 53 MHz OH line near G48.4$-$1.4 using the National MST Radar 
Facility }

\author[S. M. Menon et. al. ] 
{Srikumar M. Menon$^{1}$\thanks{E-mail:sm.menon@mit.manipal.edu},
D. Anish Roshi$^{2}$\thanks{E-mail:anish@rri.res.in}  and 
T. Rajendra Prasad$^{3}$\thanks{E-mail:trp1910@rediffmail.com}\\
$^{1}$Manipal Institute of Technology, Manipal 576 104, India \\
$^{2}$Raman Research Institute, Sadashivanagar, Bangalore 560 080, India \\
$^{3}$National MST Radar Facility, Gadanki 517 112, India \\}

\begin{document}

\date{Accepted 2004 October 12. Received 2004 October 6; in original 
form 2004 August 3}

\pagerange{\pageref{firstpage}--\pageref{lastpage}} \pubyear{}

\maketitle

\label{firstpage}

\begin{abstract}
We present the results of a search for the ground state hyperfine 
transition of the OH radical near 53 MHz using the National MST Radar 
Facility at Gadanki, India. The observed position was G48.4$-$1.4 near 
the Galactic plane. The OH line is not detected. We place a 3$\sigma$ 
upper limit for the line flux density at 39 Jy from our observations.
We also did not detect recombination lines (RLs) of carbon, which were 
within the frequency range of our observations. The 3$\sigma$ upper 
limit of 20 Jy obtained for the flux density of carbon RL, along with
observations at 34.5 and 327 MHz are used to constrain the physical
properties of the line forming region. Our upper limit is consistent 
with the line emission expected from a partially ionized region with 
electron temperature, density and path lengths in the range 
20 -- 300 K, 0.03 -- 0.3 cm$^{-3}$ 
and 0.1 -- 170 pc respectively (Kantharia \& Anantharamaiah 2001).
\end{abstract} 


\section{Introduction}

A large variety of molecules, including complex organic ones, have been 
detected in the galactic interstellar medium.
The hydroxyl (OH) radical is quite abundant in the galactic plane and has 
several rotational transitions that are easily observed at microwave frequencies.
These lines are found to originate from thermal processes as well as non-thermal 
processes (i.e. maser emission).
Thermal emission from OH radical was first detected in 1963 (\nocite{w63}Weinreb et al. 1963).
The thermal lines are observed from extended regions in the Galactic plane.
On the other hand, maser emission from OH radical is associated with specific 
activities in the Galaxy.
For instance, OH masers from the ground state rotational transitions with frequencies 1665 and 1667 MHz are mostly associated with star-forming regions, 1612 MHz masers are associated with evolved stars (Elitzur 1992; Reid \& Moran 1981) and the 1720 MHz masers are associated with shocked regions at the boundaries where supernova remnants interact with molecular clouds (Wardle \& Yusuf-Zadeh 2002).
Modeling the line emission provides an understanding of the physical conditions and processes that occur in the galactic interstellar medium where OH lines originate.
Despite the large amount of observational data available and considerable 
theoretical effort, a clear understanding of the pumping mechanisms that lead 
to different inversions in OH masers is lacking (Elitzur 1976, 
Cesaroni \& Walmsley 1990).
 
In addition to the microwave lines, the OH radical also has transitions in 
the meter-wave.
These lines are produced due to transitions between hyperfine levels in the
same $\lambda$ doublet states (see Section~\ref{sec:ohline}).
The frequencies of these transitions in the ground rotational state are 53 and 55 MHz.
These lines, which have not been observed so far, are expected to have weak line intensities because they are magnetic dipole transitions.
Moreover, observations are further complicated due to excessive man-made radio
frequency interference (RFI) near the line frequencies.
It is owing to severe RFI at this frequency, along with the weak nature of these lines, that attempts to detect these lines were not made earlier (Turner, B. E. personal communication).
As discussed above, in a variety of astrophysical situations, maser emission is observed from the microwave transitions.
Therefore, the possibility of maser emission of meter-wave transitions cannot be ruled out and may be strong enough to be detected.
The detection of these lines could provide clues to resolve, for example, the 
pumping mechanism of OH masers observed at microwave frequencies.

In this paper, we present an attempt to detect the 53 MHz OH line by observing with the National MST Radar Facility (NMRF) at Gadanki, India, in the receiving mode of the antenna.
A brief discussion of the 53 MHz OH line emission is given in Section~\ref{sec:ohline}.
We describe the observing setup and strategy in Section~\ref{sec:obs} and discuss the data analysis in Section~\ref{sec:dat}.
Results of the OH line observations are presented in Section~\ref{sec:res}.
In addition to the OH line, carbon recombination lines (RLs) were also present within the observing band.
The results of the RL observations are also included in Section~\ref{sec:res}. Our conclusion and prospects for future observations are given in Section~\ref{sec:con}.

\section{53 MHz OH line emission}
\label{sec:ohline}

\begin{figure}
\includegraphics[width=3.0in, height=4.0in, angle = -90]{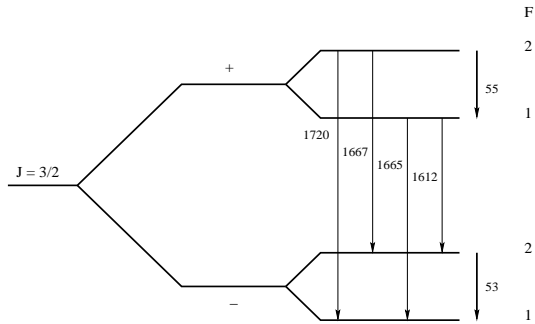}
\caption{Energy diagram of the hyperfine levels for the ground state rotational 
transitions of the OH radical. The transition frequencies are given in MHz.}
\label{fig:eng}
\end{figure}

The energy diagram for the ground state rotational transitions of the OH molecule is shown in Fig.~\ref{fig:eng}.
The rotational ladder of OH is split into $^2\Pi_{\frac{1}{2}}$ and $^2\Pi_{\frac{3}{2}}$ ladders because of the spin-orbit coupling of a single unpaired electron in a 2p orbital on the O atom.
Here we consider the ground rotational state, characterized by $J = \frac{3}{2}$.
This state is split into two levels as a result of the small difference in moment of inertia when the electron orbital is parallel or perpendicular to the molecule's rotation axis ($\lambda$ doubling) and further split by the hyperfine interaction between the electron spin and the spin of the proton in the H atom.
The microwave lines of OH at frequencies 1612, 1665, 1667 and 1720 MHz arise from transitions between these four $\lambda$ doubling states, i.e., $J = \frac{3}{2}$, $F^+ \rightarrow F^-$, where + and $-$ indicate the higher and lower $\lambda$ doublet states.  
(see Fig.~\ref{fig:eng}).
The two magnetic dipole transitions from quantum states $F = 2^+ \rightarrow 1^+$ and $F = 2^- \rightarrow 1^-$ have frequencies near 55 and 53 MHz respectively.

%To begin with, we used level transition probabilities to estimate the critical density as well as the thermal line strength.
%We used a value of $10^{-18}$ for the Einstein A coefficient for the 53MHz transition.
%This followed from a statement by Weaver (1963) that A for this transition is approximately $10^{-7}$ times the A value for the 1.6GHz electric dipole transitions of OH, which is $10^{-11}$ (Destombes et al. 1977).

We estimate the thermal line intensity from a cloud with OH column density
in the range 10$^{14}$ to 10$^{17}$ cm$^{-2}$ and a typical kinetic temperature of 100 K
(Weinreb et al. 1963, Elitzur 1976, Lockett et al. 1999). The line optical depth
is estimated using the Einstein A coefficient for the 53 MHz transition of 
$\sim$ 10$^{-18}$ s$^{-1}$, the rotational constant of 18.51 cm$^{-1}$  and 
considering a typical OH line width of 5 \kms 
(Weaver 1963, Destombes et al. 1977, Turner 1979). A mean galactic background temperature
of 25000 K at 53 MHz is used for the calculation. This background temperature 
is obtained by scaling the measured temperature at 34.5 MHz using a spectral
index of $-$2.7 (Dwarakanath \& Udaya Shankar 1990, \nocite{sb88}Salter \& Brown 1988)
The expected line brightness temperature is 10$^{-3}$ and 1 K
for column densities 10$^{14}$ and 10$^{17}$ cm$^{-2}$ respectively. 
Due to the high galactic background
near 53 MHz (which dominates the system temperature) it is not possible to detect the thermal
OH line in a reasonable observing time. However, there exists a possibility of maser emission 
of the meter-wave transitions (Turner, B. E. personal communication).
To our knowledge, there have been no attempts to calculate the line strengths of 
these maser lines. We decided to search for 53 MHz maser emission
towards direction where OH microwave masers (as well as thermal lines) were detected earlier.
 
\section{Observations}
\label{sec:obs}

The observations presented in this paper are made with the National MST Radar Facility located at Gadanki in Andhra Pradesh, India (13\degree $27'$ $21."12 N$, 79\degree $10'$ $37."10 E$).
In this section, we briefly describe the radar facility, the back-end used for the observations and the selection of sources for observation.

\subsection{The MST Radar Facility}
\label{sec:obs:mst}

\begin{figure}
\includegraphics[width=3.0in, height=4.0in, angle = -90]{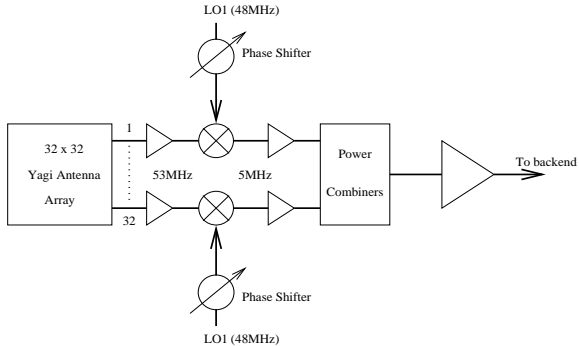}
\caption{Block diagram of the receiver system of the National MST Radar Facility.}
\label{fig:mst}
\end{figure}

A simplified block diagram of the receiver system of the MST Radar Facility is 
shown in Fig.~\ref{fig:mst} (see Rao \etal\ 1995 for details).
The antenna used for transmission and reception of radar signals is a phased array 
with 1024 crossed Yagi elements, placed in a 32 x 32 matrix.
The array is spread out over an area of 130m x 130m and has an effective aperture of $10^4$ m$^2$.
The antenna and receiver system are tuned to operate at a center frequency of 53 MHz 
with a bandwidth of $\sim$ 2 MHz.
The two crossed Yagi elements, corresponding to the two orthogonal linear polarizations, 
are oriented along the East-West (E-W) and North-South (N-S) directions.
The signals from the elements are combined to form beams that can be pointed 
in steps of 1\degree\ to a maximum of 20\degree\ from the zenith along E-W and N-S directions.
The E-W beams are formed in the following way.
The signals of 32 N-S Yagi elements, which form a linear sub-array, are combined 
using passive directional couplers. Such a sub-array is also called a module.
The output of 32 such modules are amplified using low-noise amplifiers and 
down converted to 5 MHz intermediate frequency (IF) using a local oscillator 
(LO) of frequency 48 MHz.
These IF outputs are combined with appropriate phase gradients to form the E-W beams.
The phase-shifters are placed in the LO path as shown in Fig.~\ref{fig:mst}.
Signals from the Yagi elements oriented in the E-W direction are combined in a 
similar fashion to form the N-S beams.
The half power beam width of the antenna is $\sim$3\degree.
The E-W beams can be pointed in steps of 1\degree\ to a maximum of 20\degree\ from the zenith.
For our observations, we used the different E-W beams to track the celestial source.
Note that the MST array is oriented to the geomagnetic meridian which is 
oriented 2\degree due West with reference to the geographic meridian at Gadanki.
This orientation of the array along with the location of the facility at a 
latitude of $\sim$13\degree.5 makes the source move away from the beam centre 
while tracking the source using the E-W beams (see Section.~\ref{sec:obs:strat} for
further discussion).

\subsection{The back-end}
\begin{figure}
\includegraphics[width=3.0in, height=3.5in, angle = -90]{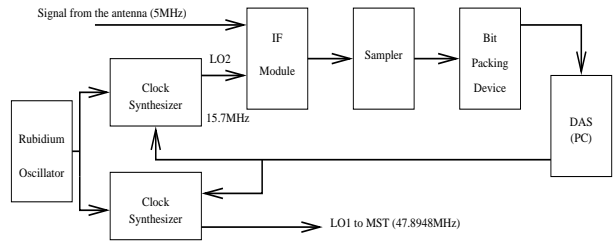}
\caption{Block diagram of the back-end used for observations. }
\label{fig:ppr}
\end{figure}

The Portable Pulsar Receiver (PPR; Deshpande et al. 2004) developed at the Raman 
Research Institute was used as the backend for the observations.
The block diagram of the PPR is shown in Fig.~\ref{fig:ppr}.
The 5 MHz IF from the radar receiver is mixed with a local oscillator of frequency 15.7 MHz to obtain a second IF at 10.7 MHz.
The local oscillator is generated using a synthesizer in the PPR and its frequency can be programmed from the computer used for data acquisition.
The bandwidth of the second IF is restricted to $\sim$1.3 MHz using a bandpass filter centred at 10.7 MHz.
The IF signal is bandpass sampled at 3.3 MHz and converted to 2-bit, 4-level data. A bit-packing logic then packs consecutive samples into 16-bit words.
This data is acquired and stored using a personal computer. 
A marker is also inserted after 4096 data words to identify any loss of samples while analysing data.
The PPR has another synthesizer which we used to generate the first local oscillator signal (see below).
Both the synthesizers are locked to a rubidium oscillator.

 \subsection{Modification of PPR data acquisition software}
The existing data acquisition software for the PPR was modified to incorporate a dual-Dicke frequency switching scheme for bandpass calibration.
We switched the first local oscillator (LO1) between 47.8948 MHz and 48.0 MHz.
The values of the frequencies were selected so that the 53 MHz line of OH and five RL of carbon be within the $\sim$1.3 MHz bandwidth used for observations.
The difference between the two frequencies was chosen based on the expected line width of the RL.
The modified data acquisition software sets the first local oscillator frequency and acquires 640 KBytes of data corresponding to each frequency setting after allowing for settling time of a few msec.
This amount of data will be acquired in about a sec.
Thus the frequency switching takes place at 1 sec rate.

\subsection{Observing Strategy and Source Selection}
\label{sec:obs:strat}

The candidate sources towards which the 53 MHz OH line emission 
is expected are mostly in the Galactic plane.
The long integration required for our observations can be obtained 
if the antenna could track the celestial source.
However, the MST radar facility in its current configuration does 
not permit tracking of celestial sources.
We adopted a strategy of observing a position in the Galactic plane 
with the different East-West beams of the antenna, thus effectively 
tracking the position for about 40\degree (see Section~\ref{sec:obs:mst}).
This position corresponds to the galactic co-ordinates 
$l = $ 48\degree.4 and $b = -1$\degree.4 (RA(2000) = 19h25m33.0s, 
Dec(2000) = $+13$\degree\ 06$^{'}$ 35.2$^{''}$). Note that G48.4$-$1.4 
is the only position in the inner Galaxy that can be tracked for 40\degree
in this mode using the present configuration of the MST radar.

Microwave masers of OH have been detected within the $effective$ beam
area ($\sim$ 4\degree; see below) of the observation centered at G48.4$-$1.4
(see Table~\ref{tab1}). Most of these masers are associated with the star 
forming region W51 (Gaume \& Mutel 1987). All the four OH lines were detected
from this region and typically the main lines (1665 \& 1667 MHz) are seen 
to be bright in emission. Anomalous 1720 MHz OH maser emission associated with 
shocked molecular gas adjacent to the supernova remnant G49.2$-$0.7 was also
present within the beam area (Frail \etal 1996). Moreover, Roshi \& 
Anantharamaiah (2000) have detected carbon RLs towards G50+0.0 with a
beam of 2\degree. Thus we decided to observe the direction G48.4$-$1.4
with the MST radar facility. 

The direction G48.4$-$1.4 was observed in a total of 13 East-West beams separated 
by the half-power beam-width of 3\degree, resulting in an integration 
time of approximately 2.5 hours on each observing run.
As discussed in Section~\ref{sec:obs:mst}, in this mode of observation, 
the sky position shifts from the beam center by about $\pm0$\degree.4 in 
RA and $\pm0$\degree.75 in Dec. Thus the effective area observed will 
be 3\degree.8  $\times$ 4\degree.5 centered at G48.4$-$1.4.

As discussed earlier RFI is a major factor in determining the success of observations 
at this frequency.
To minimize man-made interference, observations were to be made during night time.
G48.4$-$1.4 could be observed at night with the MST radar during July-August.
We conducted our observations from 30 July to 02 August 2003 between 21:30 
and 00:30 hours local time.

While observing, we radiated a tone at frequency 52.6 MHz to make sure that the 
receiver and data acquisition system were functioning as expected and to confirm that 
the frequency switching was taking place as desired. The radiated frequency was chosen 
to lie outside the frequency range where the OH line and the RLs are expected.

\begin{table}
\centering
\begin{minipage}{7in}
\caption{OH maser and RL sources within 4\degree\ of G48.4$-$1.4.}
\label{tab1}
\begin{tabular}{lll}
\hline
Source & Type & Reference    \\ \hline
W51   & OH Maser & Gaume \& Mutel (1987)    \\
G49.2$-$0.7 & Anomalous 1720 MHz  & Frail et al. (1996)    \\
            &  maser              &  \\
G50$+$0.0 & Carbon RL & Roshi \& \\
          &           & Anantharamaiah (2000)    \\     \hline
\end{tabular}
\end{minipage}
\end{table}

\section{Data Analysis}
\label{sec:dat}

We developed software in C/C++ and AIPS++ (Astronomical Image Processing
System) to reduce the data. The data analysis starts with the extraction of 
the marker, which is inserted along with the data, to check for any loss of data.
If a loss of data was detected, the 2 $\times$ 640 KBytes of data 
corresponding to the two frequency settings were discarded.
The 2-bit quantized data are then assigned voltage values of $-$3, $-$1, $+$1 or $+$3
as recommended from the hardware specification of the PPR.
A 4096 point FFT of these voltages is taken to obtain a 2048 point power spectrum.
The frequency resolution of the spectrum is 807.5 Hz which corresponds to $\sim$
4.6 kms$^{-1}$.
These power spectra were averaged for about 1 sec.
The spectra obtained when LO1 was set to 47.8948 MHz ($S_{LO1}$) and those obtained with 48.0 MHz ($S_{LO2}$) were averaged separately.
Both averaged spectra were carefully examined for interference.
Spectra badly affected by interference were discarded.
Spectra showing interference in a few channels were edited using a channel weighting scheme.
In this scheme, the channels affected by interference were given a weight of zero (Roshi \& Anantharamaiah 2000).
The bandpass calibrated spectra were obtained as $\frac{S_{LO1} - S_{LO2}}{S_{LO2}}$.
The calibrated spectrum along with the channel weights were weighted averaged to obtain the final spectrum.
This method resulted in slightly different integration times for each channel. 
The maximum difference in integration time in the final spectrum was less than 10 percent.

An average spectrum  for each day's observation was obtained without shifting 
the spectra in velocity since the correction for the antenna motion relative
to the Local Standard of Rest (LSR) during one observing run ($\sim$ 2.5  hours) 
was less than the spectral resolution. In the frequency switching scheme 
(double-Dicke frequency switching) used for observations, the expected lines 
are present at two spectral ranges in the calibrated spectrum separated by 105.2 kHz.
The average spectrum is obtained by folding these two frequency ranges.
The four such spectra obtained from each day's observing run were then 
averaged after applying the required correction for the LSR motion of the antenna
to get the final spectrum.

During data analysis, we found that most of the interference was internal to the MST radar.
These interference were caused by oscillations from some of the front-end amplifiers.
On some occasions, the interference was so pronounced that some of the modules of
the radar antenna had to be turned off. There was some interference due to 
lightning on most days. External man made RFI was seen to be very low.

\section{Results}
\label{sec:res}

The final spectra are in units of $\frac{T_L}{T_{sys}}$,
where $T_L$ is the line antenna temperature 
and $T_{sys}$ is the system temperature.
The system temperature is estimated as follows.
The sky background radiation (= 26930 K) 
towards the observed position is obtained after scaling the 
measured values at 34.5 MHz and using a spectral index of $-$2.7
(Dwarakanath \& Udaya Shankar 1990, Salter \& Brown 1988).
The measured receiver temperature of the radar facility 
is $\sim$ 4630K (Damle et al. 1992).
Thus the total system temperature is $\sim$ 31560 K. 
During our observations interference due to lighting was
present, which would have increased the system temperature. 
However, this increase in system temperature is not significant 
since the duration of data affected by lightning 
is less than a few percent of the total observing time.
The line antenna temperatures obtained from the spectra
are converted to units of flux density $S_{L}$ using the equation 
$\frac{1}{2} S_L A_{eff} = k T_L$, where $A_{eff}$ is the 
effective area of the antenna. Taking into account 
the fact that some of the modules were turned off during the 
observations (see Section~\ref{sec:dat}), we estimated $A_{eff}$ is $\sim$ 9060 m$^2$.
This effective area is consistent with that obtained from
the calibration observation done on Virgo A (3C274).

%The equivalent flux obtained is 9616.8Jy which is used to 
%convert the spectral values to units of flux density.

\subsection{53 MHz OH line}

\begin{figure}
\includegraphics[width=3.0in, height=3.5in, angle = -90]{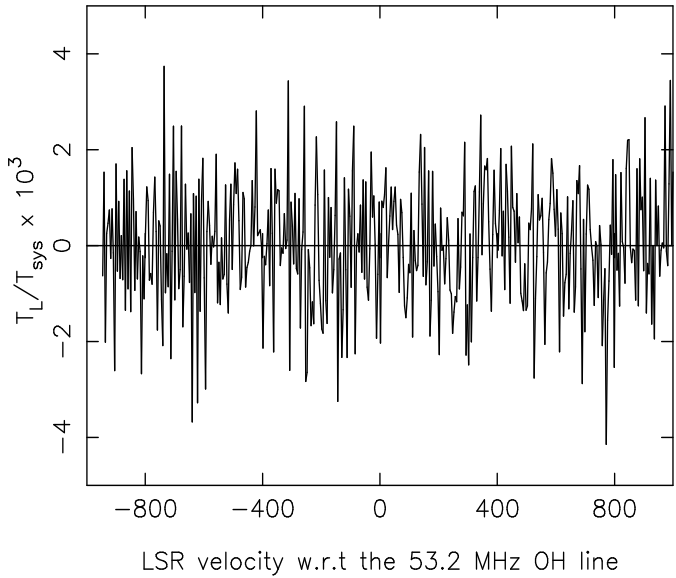}
\caption{Spectrum near 53.16 MHz obtained towards G48.4$-$1.4.  
OH line was not detected to a 3$\sigma$ level 
of 4 $\times 10^{-3}$ in units of $\frac{T_L}{T_{sys}}$,
which corresponds to an upper limit on flux density of 39 Jy.}
\label{fig:ohspec}
\end{figure}

The rest frequency of the OH line is 53.1595 MHz.
The final spectrum obtained near this frequency range is 
shown in Fig.~\ref{fig:ohspec}. No line was detected to a 3$\sigma$ level 
of 4 $\times 10^{-3}$ in units of $\frac{T_L}{T_{sys}}$.
This corresponds to an upper limit on flux of 39 Jy.
Assuming a typical angular size of $\sim$ 100 mas for OH maser
emission (Hoffman et al. 2003), the brightness temperature 
of 53 MHz OH maser line is $<$ 2 $\times$ 10$^{12}$ K. 
Comparing this temperature with the estimated thermal line temperature
(see Section~\ref{sec:ohline}), we rule out maser amplification factor $>$ 
a few times 10$^{12}$.

\subsection{Carbon Recombination Lines}

\begin{figure}
\includegraphics[width=3.0in, height=3.5in, angle = -90]{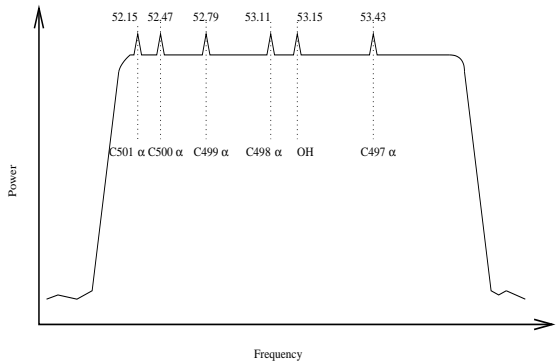}
\caption{Schematic representation of the positioning of the OH line and 
carbon RLs within the observed bandwidth.}
\label{fig:freqs}
\end{figure}

\begin{figure}
\includegraphics[width=3.0in, height=3.5in, angle = -90]{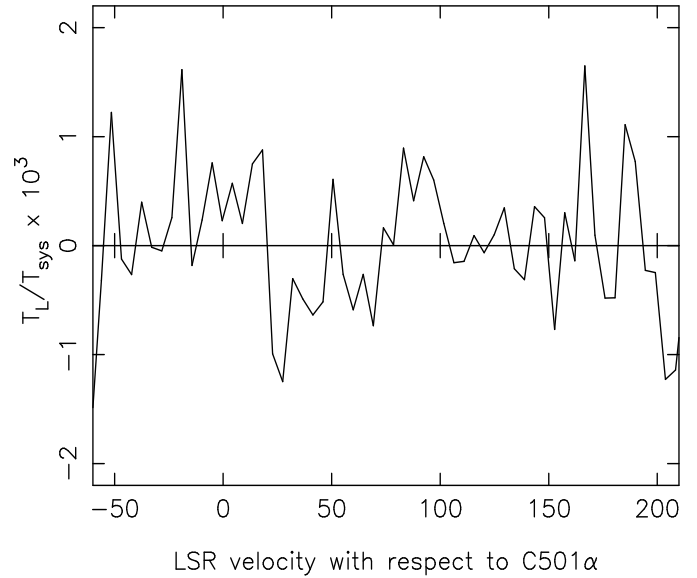}
\caption{Spectrum obtained towards G48.4$-$1.4 by averaging 
the spectra corresponding to five carbon RL transitions 
near 53 MHz. The observed RL transitions are C501$\alpha$(52.1576 MHz), 
C500$\alpha$ (52.4709 MHz), C499$\alpha$ (52.7867 MHz), 
C498$\alpha$ (53.1050 MHz) and C497$\alpha$ (53.4259 MHz). Carbon
RL was not detected to a 3$\sigma$ upper limit of 2 $\times$ 10$^{-3}$
in units of $\frac{T_L}{T_{sys}}$, which corresponds to an upper limit 
on the flux density of 20 Jy.}
\label{fig:rrl}
\end{figure}

Five recombination lines of carbon are present within the observed bandwidth 
(see Fig.~\ref{fig:freqs}). They are the C501$\alpha$(52.1576  MHz), 
C500$\alpha$ (52.4709 MHz), 
C499$\alpha$ (52.7867 MHz), C498$\alpha$ (53.1050 MHz) and C497$\alpha$ (53.4259 MHz) 
transitions. To improve the signal to noise ratio, we averaged the spectral 
values over the frequency range where these lines are expected.
The final spectrum thus obtained is shown in Fig.~\ref{fig:rrl}.
No line feature was detected to a 3 $\sigma$ level of 2 $\times$ $10^{-3}$ in 
units of $\frac{T_L}{T_{sys}}$, which corresponds to an upper limit on the 
flux density of 20 Jy.

Extensive surveys of carbon RLs from the galactic plane have been made earlier
near 34.5 and 327 MHz (Kantharia \& Anantharamaiah 2001, Roshi \& Anantharamaiah 2000).
These observations were used to determine the physical properties of the 
carbon line forming region. Partially ionized regions with electron temperature
between 20 -- 300 K, electron densities in the range 0.03 -- 0.3 cm$^{-3}$ and
path lengths between 0.1 -- 170 pc can produce the observed carbon RL intensity
at different frequencies (Kantharia \& Anantharamaiah 2001). The observations at 53 MHz 
towards G48.4$-$1.4 along with data at other frequencies can be used to constrain 
the physical properties of the carbon line forming region in this direction. 
Carbon RL was detected at 327 MHz towards G50.0$+$0.0 with a 2\degree\ beam. 
Line was not detected towards G47.8+0.0 at 327 MHz
implying that the line forming region is less than 4\degree\ in extent. The
observation towards G50.0$+$0.0 has about 1\degree\ overlap with the region
observed at 53 MHz. No RL was detected at 34.5 MHz towards G50.0+0.0. This
observation has an angular resolution of 21\arcmin $\times$ 25\degree. Combining
these data, we found that carbon line forming region with physical properties 
in the range determined by Kantharia \& Anantharamaiah (2001) is consistent 
with all the three observations. Observations with similar angular resolutions
are needed to better constrain the physical properties.

\section{Conclusion and Future Possibilities}
\label{sec:con}

In this paper, we have presented observations made near 53 MHz with the 
aim of detecting the 53.2 MHz transition of the OH radical.
This OH line is due to the transition between the hyperfine levels
in the same $\lambda$ doublet state of lowest energy.
The observations were conducted using the National MST Radar Facility at Gadanki. 
No OH line was detected to a 3$\sigma$ upper limit of 39 Jy.
Within the observed bandwidth, five carbon RLs were present.
We have not detected the carbon lines to a 3$\sigma$ limit of 20 Jy.

External man-made interference is low at Gadanki, which is very 
encouraging for future attempts to search for the OH line in other positions 
in the Galactic plane. However, currently, a major limitation is the lack of 
tracking facility for the MST radar antenna. A tracking facility would 
allow us to search for OH line at other positions in the Galactic plane 
in a reasonable observing time.

\section*{Acknowledgments}

We gratefully acknowledge the help of S. Gurushant of Raman Research Institute 
for helping us set up the PPR as well as for his help with the observations. We thank
Prof. N. Udaya Shankar for extending the help of the Radio Astronomy
Lab at RRI for this project.
We are grateful to T. Prabu, C. Vinutha, B. S. Girish and H. N. Nagaraj of Raman 
Research Institute, Bangalore for their valuable help in setting up the PPR for 
our observations.
The PPR was developed by Prof. A. A. Deshpande and team at RRI.
We are thankful for discussions with Prof. A. A. Deshpande that led to the 
initiation of this project.
We are indebted to G. Sarabagopalan of RRI for his generous help in contacting NMRF and in planning the observations.
We are extremely grateful to Prof. D. Narayana Rao, Director, NMRF for speedy approval of our observing proposal.
His enthusiastic support was instrumental in the efficient conduct of this project.
It is a pleasure to thank S. Kalesha, S. Babu and all the staff at NMRF, Gadanki for their help with the observations and a memorable stay at Gadanki.
SMM and DAR are grateful to NMRF for providing the observing time at the MST Radar and hospitality. 
SMM wishes to thank RRI for hospitality as well as use of resources during the conduct of this project.
He also wishes to thank Sudhakara G., R. Samanth, M. Raghuprem and P. C. Madhuraj of Manipal Institute of Technology, Manipal for taking over some academic responsibilities so that the entire vacation could be devoted to this project. We thank the anonymous referee for
his helpful suggestions for the continuation of the work presented in the paper.

\label{lastpage}

\end{document}